\documentclass[
superscriptaddress,
preprintnumbers,
nofootinbib,
nobibnotes,
notitlepage,
amsmath,amssymb,
aps,
twocolumn,
prl,
]{revtex4-2}

\usepackage{textcomp}
\DeclareUnicodeCharacter{2212}{\textminus}

\usepackage{hyperref}
\usepackage{xcolor}
\usepackage{graphicx}
\usepackage{enumitem}
\usepackage{ulem}
\setlist[description]{
    font=\normalfont\itshape, 
    leftmargin=0pt,          
    labelindent=0pt,          
    topsep=0pt,itemsep=0pt,parsep=0pt
}

\newcommand{\summnu}{\sum m_\nu}
\newcommand{\neff}{N_\mathrm{eff}}
\newcommand{\dneff}{\Delta N_\mathrm{eff}}
\newcommand{\nsp}{N_{\nu_{\rm s},\phi}}
\newcommand{\ms}{m_{\nu_{\rm s}}}

\begin{document}

\preprint{TTK-26-15}

\title{Recoupled Dark Radiation reconciling CMB and DESI BAO measurements}

\author{Ravi Kumar Sharma}
\email{rksharma@physik.rwth-aachen.de}
\affiliation{Institute for Theoretical Particle Physics and Cosmology  (TTK),
RWTH Aachen University, Sommerfeldstr. 16, D-52056 Aachen, Germany}

\author{Maria Archidiacono}
\email{maria.archidiacono@unimi.it}
\affiliation{Dipartimento di Fisica ``Aldo Pontremoli'', Università degli Studi di Milano, Via Celoria 16, 20133 Milano, Italy}
\affiliation{INFN, Sezione di Milano, Via Celoria 16, 20133 Milano, Italy}

\author{Julien Lesgourgues}
\affiliation{Institute for Theoretical Particle Physics and Cosmology  (TTK),
RWTH Aachen University, Sommerfeldstr. 16, D-52056 Aachen, Germany}


\date{\today}

\begin{abstract}
Recent DESI BAO measurements, combined with CMB data, reveal a tension within the $\Lambda$CDM model that leads to a discrepancy between cosmological and laboratory bounds on the summed neutrino mass.
We show that a recoupled interacting radiation component can alleviate this cosmological tension, as well as the one with neutrino oscillation experiments.
Sterile neutrinos interacting through a light pseudoscalar mediator provide a concrete realization of this scenario.
The resulting interacting fluid modifies the CMB phenomenology, lowers the preferred matter density, and improves the consistency between CMB and DESI BAO measurements.
Combining CMB with DESI DR2 BAO measurements, we find a $2.7 \sigma$ preference for a nonzero interacting sterile neutrino component, $\nsp=0.253 \pm 0.094$, corresponding to an improvement $\Delta \chi^2=-8.98$ relative to $\Lambda$CDM. The model also reduces the tension with the SH0ES determination of the Hubble constant to the $2.4\sigma$ level. 
\end{abstract}


\maketitle

\section{Introduction}
\label{sec:introduction}

The emerging tension between Cosmic Microwave Background (CMB) and DESI baryon acoustic oscillation (BAO) data \cite{DESI:2025zgx} poses a challenge to the $\Lambda$CDM model.
The discrepancy originates from the mismatch between CMB and DESI BAO measurements in the fractional matter density parameter $\Omega_{\rm m}$ and in the combination of the Hubble constant and the sound horizon at baryon drag $H_0r_{\rm d}$ (see Ref.~\cite{Loverde:2024nfi, Graham:2025dqn}).
A striking consequence of this tension is the preference for a very small or vanishing neutrino mass, below the minimum value allowed by neutrino oscillation experiments in normal ordering \cite{DESI:2025ffm}. A cosmological ingredient with effects opposite to that of massive neutrinos would even be preferred - a situation often referred to as a preference for `negative neutrino mass' \cite{Elbers:2025vlz}. 
While we wait for ongoing and future LSS surveys, such as Euclid \cite{Euclid:2024yrr} and the Vera Rubin Observatory \cite{LSSTDarkEnergyScience:2018jkl}, which are expected to improve our understanding of neutrino properties \cite{Euclid:2024imf, Racco:2024lbu}, several models have been proposed in order to solve this discrepancy, including an increased optical depth to reionization \cite{Sailer:2025lxj}, nonzero curvature \cite{Chen:2025mlf}, a variation of the lensing amplitude \cite{Cozzumbo:2025ewt}, or early dark energy \cite{Chaussidon:2025npr}.

The emerging discrepancy may instead point toward nonstandard radiation dynamics around recombination. In particular, a relativistic component that recouples before recombination and subsequently annihilates into light degrees of freedom can modify the CMB phenomenology while avoiding the late-time suppression of structure associated with massive free-streaming species. 
One concrete realization of this mechanism is provided by sterile neutrinos with pseudoscalar self-interactions, as initially proposed in Ref.~\cite{Archidiacono:2014nda}.
Sterile neutrinos in the eV mass range have long been invoked to explain the oscillation anomalies observed in short-baseline experiments following the original LSND claim \cite{PhysRevD.64.112007}.
Despite the tension present in global analyses of laboratory bounds \cite{Giunti:2026uaf, Lister:2026jab, MicroBooNE:2025nll, KATRIN:2025lph}, the current status of sterile neutrinos is still debated.
An extensive experimental program is underway to provide a definitive answer about their existence \cite{Diaz:2019fwt}.
The simplest sterile neutrino models are excluded by cosmological data because the mixing angle preferred by oscillations experiments would fully thermalise them in the Early Universe, yielding an additional contribution $\dneff \sim 1$, and therefore $\neff \sim 4$ \cite{Gariazzo:2019gyi}.
However, nonstandard neutrino interactions (NSI) mediated by either a massive vector boson \cite{Hannestad:2013ana, Dasgupta:2013zpn, Chu:2018gxk} or by a light pseudoscalar \cite{Archidiacono:2014nda} can suppress thermalization and thus reconcile sterile neutrinos with cosmology. Additionally, pseudoscalar mediated interactions can also alleviate the Hubble tension \cite{Archidiacono:2020yey}. 
Similar effects may arise from self-interactions in the active neutrino sector mediated by heavy \cite{Cyr-Racine:2013jua, Kreisch:2019yzn, Brinckmann:2020bcn, Das:2020xke} or light \cite{Archidiacono:2013dua, Forastieri:2019cuf} particles, although these scenarios appear increasingly disfavoured by cosmological data \cite{Camarena:2024daj, AtacamaCosmologyTelescope:2025nti, Poudou:2025qcx}. 
An additional advantage of interactions confined to the sterile sector is that they can evade several astrophysical constraints from supernovae \cite{Farzan:2002wx, Das:2017iuj, Shalgar:2019rqe, Chang:2022aas, Ivanez-Ballesteros:2024nws} and high-energy neutrinos \cite{Ioka:2014kca, Bustamante:2020mep, Esteban:2021tub}, as well as laboratory bounds \cite{Brune:2018sab, Blinov:2019gcj,  Lyu:2020lps, Brdar:2020nbj} from neutrinoless double-beta decay \cite{Agostini:2015nwa, Blum:2018ljv} and meson decays \cite{Pasquini:2015fjv, Dev:2024ygx}.\footnote{This is partially true also in the case of flavour-specific NSI confined to $\tau$ neutrinos \cite{Das:2020xke, Blinov:2019gcj}.}

In this paper, we reassess the consistency of sterile neutrino self-interactions mediated by a light pseudoscalar with current cosmological data, focussing on the tension between CMB and DESI BAO data and on the associated neutrino mass problem.
Note that, although we focus on the sterile neutrino realization, the cosmological signatures discussed here arise more generally from interacting radiation components that recouple and annihilate near recombination.


\section{The pseudoscalar model}
\label{sec:model}

The coupling between the mostly sterile neutrino mass eigenstate $\nu_4$ and the pseudoscalar $\phi$ is described by the Lagrangian:
\begin{equation*}
    \mathcal{L} \sim g \, \phi \, \bar{\nu_{4}} \, \gamma_5 \, \nu_4 ,
\end{equation*}
where $g$ is the coupling constant. For simplicity, we will assume that the mediator mass $m_\phi$ is so small that $\phi$ bosons are still ultra-relativistic today. We will later refer to the sterile neutrino mass as $m_{\nu_s}=m_{\nu,4}$.
\begin{figure}[!h]
    \centering
    \includegraphics[width=\linewidth]{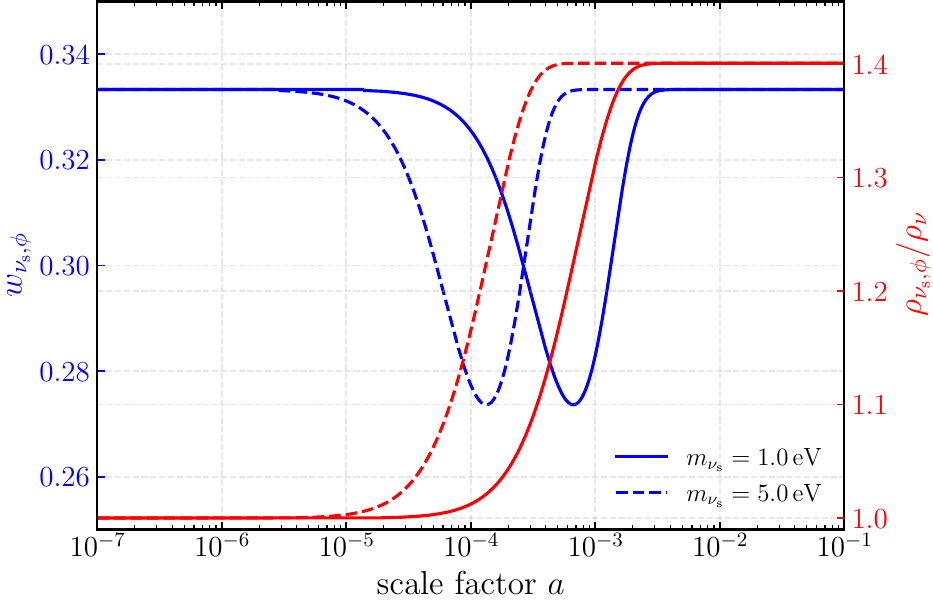}
    \caption{Evolution of the equation of state (blue) and of the background density with respect to one standard neutrino (red) of the sterile neutrino plus pseudoscalar fluid.}
    \label{fig:w_pseudo}
\end{figure}
The thermal history of the sterile neutrino plus pseudoscalar system proceeds through the following steps:
\begin{description}
\item[Partial thermalisation] in the absence of NSI, sterile neutrinos are produced through oscillations with active neutrinos around $T \sim 10$ MeV, well before neutrino decoupling at $T \sim 1$ MeV. The new interaction induces an effective matter potential $V_s \sim 0.05 g^2 T_{\nu, 4}$ \cite{Babu:1991at}, which suppresses the in-medium mixing angle and delays sterile neutrino production. As shown in Ref.~\cite{Archidiacono:2014nda}, couplings $g \gtrsim 3 \times 10^{-6}$ suppress the sterile neutrino contribution to $\dneff$ below $1$.
\item[Recoupling] the collisional scattering rate scales as $\Gamma_{\rm s} \propto g^4\,T$, while the Hubble rate scales as $H \propto T^2$. Thus, the ratio $\Gamma_{\rm s}/H$ increases as the Universe cools down, leading to a recoupling between sterile neutrinos and pseudoscalars before recombination.
\item[Annihilation] Once sterile neutrinos become nonrelativistic, they annihilate into pseudoscalars through $\nu_{\rm s}\bar{\nu_{\rm s}} \rightarrow \phi \phi$.\footnote{For the couplings considered here, annihilation dominates over decay \cite{Barenboim:2020vrr}.} Sterile neutrinos therefore effectively disappear from the cosmic neutrino background and no longer contribute to the small scale suppression of the matter power spectrum.
\end{description}
The thermal history leads to a distinctive phenomenology. After recoupling, sterile neutrinos and pseudoscalars behave as a single tightly coupled perfect fluid. 
The background evolution of the energy density and of the equation of state of the sterile neutrino plus pseudoscalar fluid is shown in Fig.~\ref{fig:w_pseudo}.
Initially the fluid behaves as radiation with $w_{\nu_{\rm s},\phi}=1/3$. As sterile neutrinos become nonrelativistic, $w_{\nu_{\rm s},\phi}$ decreases. Subsequently, annihilations into pseudoscalars drive $w_{\nu_{\rm s},\phi}$ back to $1/3$ and also increase the fluid energy density $\rho_{\nu_{\rm s},\phi}$ relative to that of one standard neutrino $\rho_\nu$ \cite{Beacom:2004yd, Archidiacono:2015oma}. 

\subsection*{Impact on CMB}
\label{sec:Impact on CMB}
\begin{figure}[h!]
    \centering
    \includegraphics[width=\linewidth]{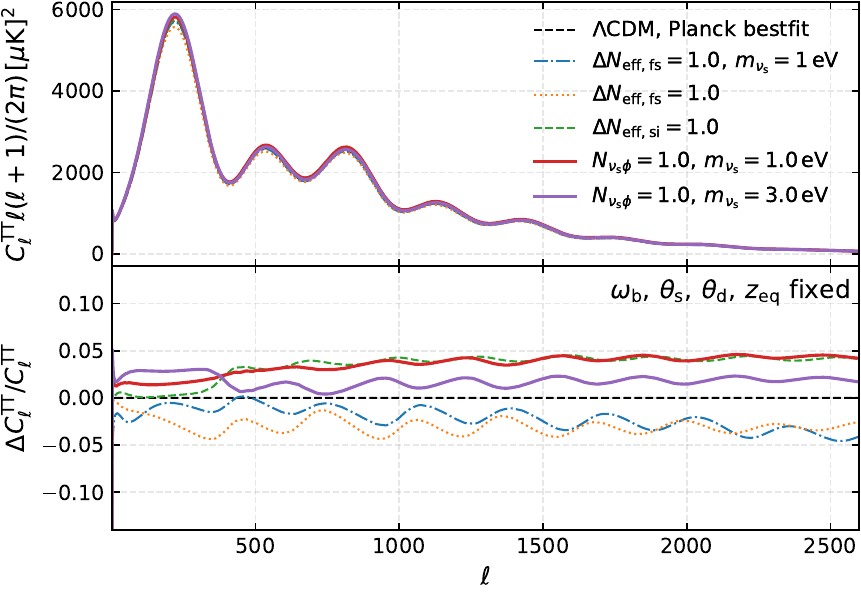}\\
    \includegraphics[width=\linewidth]{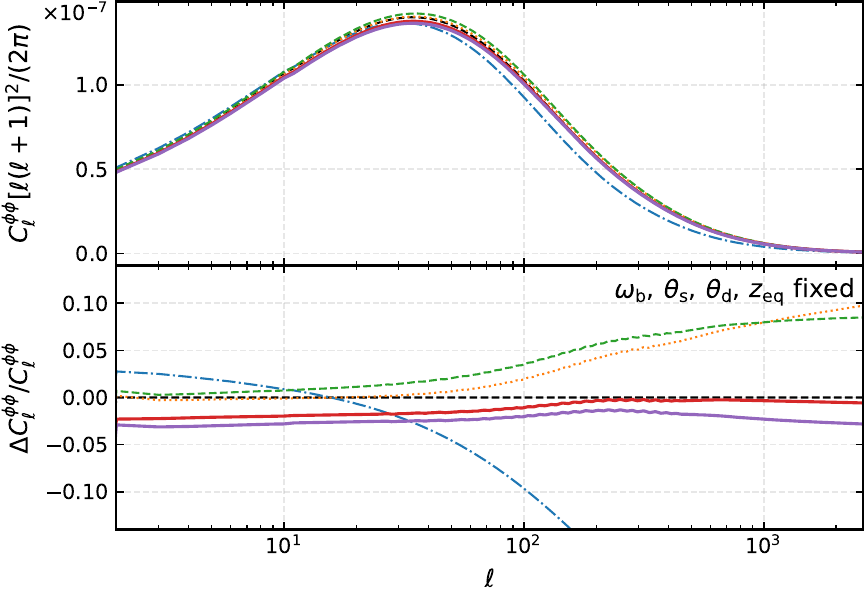}
    \caption{Lensed CMB temperature power spectrum (top figure, top panel) and lensing potential (bottom figure, top panel) for models with $\neff=4.044$ and their relative difference (bottom panels) with respect to $\Lambda$CDM with $\neff=3.044$, massless and free-streaming. The additional neutrino-like particle is free-streaming and massive (dot-dashed blue line), free-streaming and massless (dotted orange line), self-interacting (dashed green line), or interacting through a pseudoscalar with $\ms=1$ eV (solid red line) or $\ms=3$ eV (solid purple line).}
    \label{fig:ClTT}
\end{figure}

Figure~\ref{fig:ClTT} (top figure) shows the impact on the CMB temperature power spectrum of increasing $\neff$ from the Standard Model prediction $\neff = 3.044$ \citep{Mangano:2001iu,Froustey:2020mcq,Bennett:2020zkv,Drewes:2024wbw} to $\neff=4.044$.
The background effects induced by additional free-streaming light degrees of freedom, whether massive (dot dashed blue line) or massless (dotted orange line), can largely be compensated by fixing the angular size of the sound horizon at recombination $\theta_{\rm s}$ and the damping scale $\theta_{\rm d}$, which requires, respectively, increasing $H_0$ and decreasing the Helium fraction $Y_{\rm p}$ \cite{Hou:2011ec}. Thus, in figure~\ref{fig:ClTT}, we keep $(\theta_{\rm s}, \theta_{\rm d})$ fixed, as well as the baryon density $\omega_{\rm b}$ and redshift of equality $z_{\rm eq}$. However, the perturbation effects due to the modified gravitational potentials cannot be removed by adjusting background parameters \cite{Bashinsky:2003tk}. Free-streaming species induce both a scale-independent phase shift of the acoustic peaks \cite{Follin:2015hya, Baumann:2015rya, Pan:2016zla} and an additional suppression of their amplitude.

In contrast, self-interacting dark radiation (dashed green line) behaves as a fluid with negligible anisotropic stress. While the background effects are similar, the perturbation effects are different. In particular, the absence of a quadrupole moment enhances the photon monopole, increasing the overall amplitude of the CMB power spectrum and producing a scale dependent shift of the acoustic peaks.  
In the pseudoscalar scenario, the perturbation effects are closely connected to the background evolution shown in Fig.~\ref{fig:w_pseudo}. The enhancement of the acoustic peaks depends on the onset of sterile neutrino annihilations, which is controlled by the sterile neutrino mass. For $\ms=1$ eV (solid red line), the model stands very close to the case of a massless self-interacting species (dashed green line), while larger sterile neutrino masses suppress more the damping tail because the increase in the fluid energy density occurs earlier.
Figure \ref{fig:ClTT} (bottom figure) shows the corresponding effect on the CMB lensing potential. As expected, the strong suppression induced by massive sterile neutrinos (dot-dashed blue line) is erased once annihilations into pseudoscalars are included (solid lines).

\section{Methodology and data}
\label{sec:methodology}

We modified the Boltzmann solver \texttt{CLASS}\footnote{\url{https://github.com/lesgourg/class_public}}\cite{2011arXiv1104.2932L,2011JCAP...07..034B} to include a sterile neutrino plus pseudoscalar fluid. The theory predictions are then interfaced with {\sc MontePython}\footnote{\url{https://github.com/brinckmann/montepython_public}}\cite{Brinckmann:2018cvx} and emulated on the fly with \texttt{OLE}\footnote{\url{https://github.com/svenguenther/OLE}}\cite{Gunther:2025xrq}. Model diagnostic is performed with {\sc Procoli}\footnote{\url{https://github.com/tkarwal/procoli}}\cite{Karwal:2024qpt}

Our baseline $\Lambda$CDM model contains six parameters 
($H_0,\omega_{\rm b},\omega_{\rm cdm},\ln10^{10}A_{\rm s}, n_{\rm s}, \tau$).
Active neutrinos contribute as $\neff=3.044$, and their mass sum is fixed to $\sum m_{\nu}=3\times 0.02\,\mathrm{eV} = 0.06\,\mathrm{eV}$, unless otherwise stated.
The pseudoscalar model introduces two additional parameters $\nsp$ and 
$\ms$, with top-hat priors in the range $\nsp \in [0,1]$, $\ms \in [0,10] \, {\rm eV}$.  
We assume no equilibration between the active and the sterile sectors. The impact of relaxing this assumption is discussed in the Appendix.

We analyse CMB data alone and in combination with DESI DR2 BAO measurements \cite{DESI:2025zgx}. Our baseline {\it CMB} dataset includes primary anisotropies from Planck (PR4 \texttt{CamspecNPIPE} high-$\ell$ TTTEEE, PR3 low-$\ell$ TT \cite{Planck:2018vyg}, and \texttt{sroll} low-$\ell$ EE \cite{Pagano:2019tci}) and from the South Pole Telescope (\texttt{SPT-3G D1} \cite{SPT-3G:2025bzu}).\footnote{We do not include ACT primary anisotropies in our baseline CMB dataset because in models with a varying $\neff$, ACT and DESI are in tension at more than $3\sigma$ on $\Omega_{\rm m}$ \cite{SPT-3G:2025bzu}.} 
We also include CMB lensing combining Planck PR4 lensing \cite{Carron:2022eyg}, and ACT DR6 lensing  \cite{ACT:2023kun, ACT:2023dou}, and adding SPT-2YR-MUSE \cite{SPT-3G:2024atg}.

\section{Results}
\label{sec:results}
\begin{figure}
    \centering    \includegraphics[width=\linewidth]{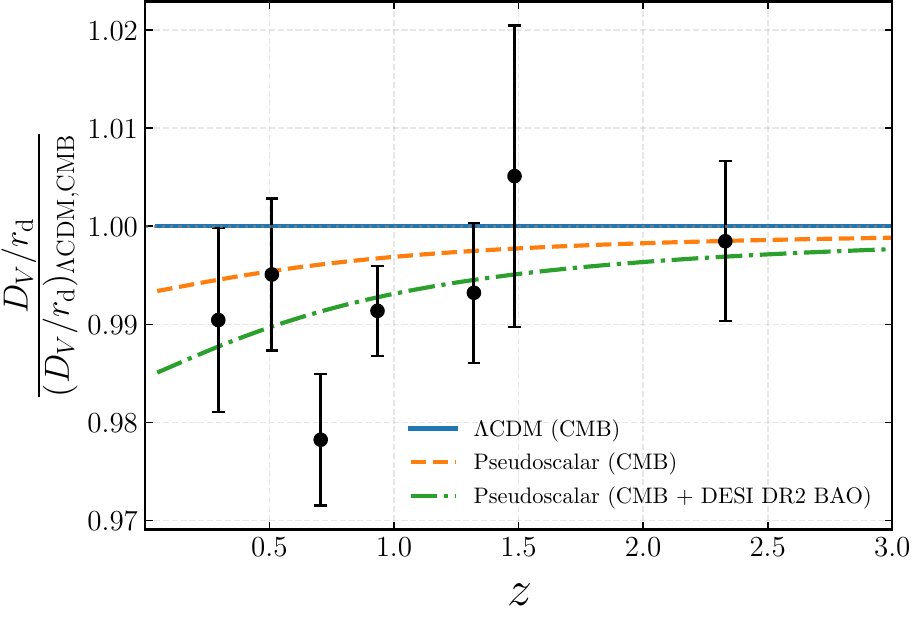}
    \caption{Bestfit prediction of the isotropic BAO distance ($D_{V}/r_{\rm d}$) in the pseudoscalar model (dashed orange line for CMB only, dot-dashed green line for CMB+DESI DR2 BAO) compared to DESI DR2 BAO measurements~\cite{DESI:2025zgx}. All data points and predictions are normalised to the CMB bestfit in $\Lambda$CDM. In all models the neutrino mass is fixed to $\summnu=0.06\,{\rm eV}$.
    }  \label{fig:bestfit_bao}
\end{figure}

Sterile neutrinos with pseudoscalar interactions provide a good fit to CMB data alone, with an overall $\chi^2$ reduced by $\Delta \chi^2=-2.86$ compared to $\Lambda$CDM.
We find $\nsp<0.40$ at the 95\% confidence level (CL), while $\ms$ is unconstrained within our prior range (see Fig.~\ref{fig:triangle} in the Appendix). In the presence of sterile neutrinos with pseudoscalar interactions, CMB data prefer a larger Hubble constant, $H_0=69.2 \pm 1.1 \, {\rm km}/{\rm s}/{\rm Mpc}$, reducing the tension with SH0ES \cite{Riess:2021jrx} to $2.6\,\sigma$. Since sterile neutrinos annihilate before structure formation, the value of $\sigma_8$ is not significantly suppressed, $\sigma_8=0.825_{-0.007}^{+0.006}$, corresponding to a shift with respect to the $\Lambda$CDM credible interval by less than $1\sigma$.
Like in \cite{Archidiacono:2020yey}, we find that the credible interval of the primordial index increases slightly to $n_s=0.972 \pm 0.006$ to compensate for the extra damping induced by a larger $\neff$ (as seen in Fig.~\ref{fig:ClTT}).\footnote{We assume that $Y_{\rm He}$ is given by the standard BBN consistency condition, therefore the damping excess cannot be compensated by varying $Y_{\rm He}$.}
\begin{figure}
    \centering    \includegraphics[width=\linewidth]{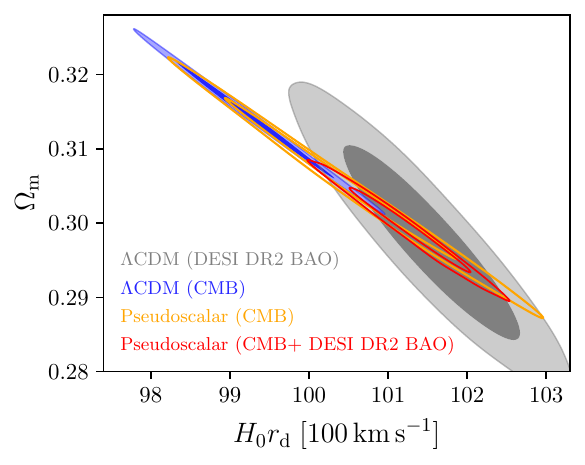}
    \caption{68\% and 95\% marginalised constraints on $\Omega_{\rm m}$ and $H_{0}\,r_{\rm d}$ in $\Lambda$CDM (blue filled contours for CMB only, grey filled contours for DESI DR2 BAO only) and in the pseudoscalar model (orange empty contours for CMB only, red empty contours for CMB + DESI DR2 BAO).}\label{fig:Omegam_H0rd}
\end{figure}
Most importantly, the pseudoscalar model favours a lower fractional matter density, $\Omega_{\rm m}$, improving the consistency between CMB and DESI BAO measurements in the $\Omega_{\rm m}$ vs $H_0\,r_{\rm d}$ plane, as shown in Fig.~\ref{fig:Omegam_H0rd}. As a consequence, in the pseudoscalar model, the CMB-only bestfit provides a good fit of the DESI BAO data, even though we are keeping the neutrino mass fixed at the minimum value allowed by oscillations experiments, $\summnu= 0.06$ eV (see Fig.~\ref{fig:bestfit_bao}). Thus, if the summed active neutrino mass was varied, the data would not push it to tiny (or even `negative') values.
 When DESI DR2 BAO data are included, the analysis yields a preference for a nonzero interacting sterile neutrino component, $\nsp=0.253 \pm 0.094$, with a mass only constrained to $\ms=5.8\pm2.2\,{\rm eV}$ within our prior range. The Hubble parameter is now $H_0=70.01 \pm 0.74 \, {\rm km}/{\rm s}/{\rm Mpc}$, reducing the SH0ES tension to $2.4\sigma$. As expected from Fig.~\ref{fig:Omegam_H0rd} and \ref{fig:bestfit_bao}, the fractional matter density further decreases to $\Omega_{\rm m}=0.2991 \pm 0.0037$, and the overall fit improves substantially relative to $\Lambda$CDM, $\Delta \chi^2=-8.98$.
 If we vary the summed active neutrino mass, we obtained $\summnu<0.10\,{\rm eV}$ at 95\% CL, consistent with oscillation experiments (see Fig.~\ref{fig:mnu} in the Appendix).
 Finally, we checked that adding Type Ia supernovae data from Pantheon+ \cite{Brout:2022vxf} does not alter our results.

\section{Discussion and Conclusions}
\label{sec:conclusions}

In this work, we have shown that a recoupled interacting radiation component can reconcile current CMB and DESI BAO measurements while removing the apparent cosmological preference for a tiny (if not `negative') summed neutrino mass.

In the specific realization considered here, sterile neutrinos interacting through a light pseudoscalar mediator provide an excellent fit to current CMB and BAO data and reduce the Hubble tension to the 2.4$\sigma$ level.
The success of the model originates from a distinctive cosmological evolution. After recoupling, the sterile neutrino plus pseudoscalar system behaves as a tightly coupled fluid rather than as free-streaming radiation, modifying the CMB acoustic structure. Subsequently, sterile neutrinos annihilate into pseudoscalars, preventing the late-time suppression of structure characteristic of light relics. This combination naturally improves the consistency between CMB and DESI BAO measurements, and removes the apparent conflict between cosmological data and the lower bound on neutrino masses implied by oscillation experiments.

The viability of the pseudoscalar model should be further tested using DESI full shape measurements \cite{DESI:2024jxi} and forthcoming large scale structure data, which we leave for future work.

Finally, although the framework considered here is motivated by sterile neutrinos, the results are more general. Any new light particle that recouples through a light mediator around recombination and subsequently annihilates into the mediator could generate a similar phenomenology, alleviating the discrepancy between CMB and DESI BAO, the neutrino mass problem, and the Hubble tension.

\section*{Acknowledgments}
We  thank Stefano Gariazzo, Carlo Giunti, Steen Hannestad, and Thomas Tram for useful discussions. We thank Nils Schöneberg  for his help with the MontePython likelihoods.  
Computational work was performed with computing resources granted by RWTH Aachen University
under project ‘rwth1971’ and ’p0021792’. RKS thanks
the Alexander von Humboldt Foundation for their support. 

\bibliography{pseudo_ref}

@article{Graham:2025dqn,
    author = "Graham, Peter W. and Green, Daniel and Meyers, Joel",
    title = "{New interpretations of the cosmological preference for a negative neutrino mass}",
    eprint = "2508.20999",
    archivePrefix = "arXiv",
    primaryClass = "astro-ph.CO",
    doi = "10.1103/1bqb-qlrj",
    journal = "Phys. Rev. D",
    volume = "113",
    number = "4",
    pages = "043514",
    year = "2026"
}

@article{Racco:2024lbu,
    author = "Racco, Davide and Zhang, Pierre and Zheng, Henry",
    title = "{Neutrino masses from large-scale structures: Future sensitivity and theory dependence}",
    eprint = "2412.04959",
    archivePrefix = "arXiv",
    primaryClass = "astro-ph.CO",
    doi = "10.1016/j.dark.2024.101803",
    journal = "Phys. Dark Univ.",
    volume = "47",
    pages = "101803",
    year = "2025"
}

@misc{LSSTDarkEnergyScience:2018jkl,
    author = "Mandelbaum, Rachel and others",
    collaboration = "LSST Dark Energy Science",
    title = "{The LSST Dark Energy Science Collaboration (DESC) Science Requirements Document}",
    eprint = "1809.01669",
    archivePrefix = "arXiv",
    primaryClass = "astro-ph.CO",
    reportNumber = "FERMILAB-PUB-18-465-A",
    doi = "10.2172/1471560",
    month = "9",
    year = "2018"
}

@article{KATRIN:2025lph,
    author = "Acharya, Himal and others",
    collaboration = "KATRIN",
    title = "{Sterile-neutrino search based on 259 days of KATRIN data}",
    eprint = "2503.18667",
    archivePrefix = "arXiv",
    primaryClass = "hep-ex",
    doi = "10.1038/s41586-025-09739-9",
    journal = "Nature",
    volume = "648",
    number = "8092",
    pages = "70--75",
    year = "2025"
}

@article{MicroBooNE:2025nll,
    author = "Abratenko, P. and others",
    collaboration = "MicroBooNE",
    title = "{Search for light sterile neutrinos with two neutrino beams at MicroBooNE}",
    eprint = "2512.07159",
    archivePrefix = "arXiv",
    primaryClass = "hep-ex",
    reportNumber = "FERMILAB-PUB-24-0865-PPD",
    doi = "10.1038/s41586-025-09757-7",
    journal = "Nature",
    volume = "648",
    number = "8092",
    pages = "64--69",
    year = "2025"
}

@inproceedings{Lister:2026jab,
    author = "Lister, Adam",
    title = "{NOvA's Current and Future Sterile Neutrino Searches}",
    booktitle = "{NuPhys2026}: {Prospect in Neutrino Physics}",
    eprint = "2602.11345",
    archivePrefix = "arXiv",
    primaryClass = "hep-ex",
    reportNumber = "FERMILAB-CONF-26-0079-PPD",
    month = "2",
    year = "2026"
}

@misc{Giunti:2026uaf,
    author = "Giunti, C. and Li, Y. F. and Zhang, R. P.",
    title = "{Revival of the Reactor Antineutrino Anomaly}",
    eprint = "2605.10353",
    archivePrefix = "arXiv",
    primaryClass = "hep-ph",
    month = "5",
    year = "2026"
}

@article{Diaz:2019fwt,
    author = {Diaz, A. and Arg{\"u}elles, C. A. and Collin, G. H. and Conrad, J. M. and Shaevitz, M. H.},
    title = "{Where Are We With Light Sterile Neutrinos?}",
    eprint = "1906.00045",
    archivePrefix = "arXiv",
    primaryClass = "hep-ex",
    doi = "10.1016/j.physrep.2020.08.005",
    journal = "Phys. Rept.",
    volume = "884",
    pages = "1--59",
    year = "2020"
}

@article{Elbers:2025vlz,
    author = "Elbers, W. and others",
    title = "{Constraints on neutrino physics from DESI DR2 BAO and DR1 full shape}",
    eprint = "2503.14744",
    archivePrefix = "arXiv",
    primaryClass = "astro-ph.CO",
    reportNumber = "FERMILAB-PUB-25-0168-PPD",
    doi = "10.1103/w9pk-xsk7",
    journal = "Phys. Rev. D",
    volume = "112",
    number = "8",
    pages = "083513",
    year = "2025"
}

@article{Loverde:2024nfi,
    author = "Loverde, Marilena and Weiner, Zachary J.",
    title = "{Massive neutrinos and cosmic composition}",
    eprint = "2410.00090",
    archivePrefix = "arXiv",
    primaryClass = "astro-ph.CO",
    doi = "10.1088/1475-7516/2024/12/048",
    journal = "JCAP",
    volume = "12",
    pages = "048",
    year = "2024"
}

@article{Chen:2025mlf,
    author = "Chen, Shi-Fan and Zaldarriaga, Matias",
    title = "{It's all Ok: curvature in light of BAO from DESI DR2}",
    eprint = "2505.00659",
    archivePrefix = "arXiv",
    primaryClass = "astro-ph.CO",
    doi = "10.1088/1475-7516/2025/08/014",
    journal = "JCAP",
    volume = "08",
    pages = "014",
    year = "2025"
}

@article{Sailer:2025lxj,
    author = "Sailer, Noah and Farren, Gerrit S. and Ferraro, Simone and White, Martin",
    title = "{Addressing Tensions in {\ensuremath{\Lambda}}CDM Cosmology by an Increase in the Optical Depth to Reionization}",
    eprint = "2504.16932",
    archivePrefix = "arXiv",
    primaryClass = "astro-ph.CO",
    doi = "10.1103/6r54-8lv4",
    journal = "Phys. Rev. Lett.",
    volume = "136",
    number = "8",
    pages = "081002",
    year = "2026"
}

@misc{Cozzumbo:2025ewt,
    author = "Cozzumbo, Andrea and Atzori Corona, Mattia and Murgia, Riccardo and Archidiacono, Maria and Cadeddu, Matteo",
    title = "{A short blanket for cosmology: the CMB lensing anomaly behind the preference for a negative neutrino mass}",
    eprint = "2511.01967",
    archivePrefix = "arXiv",
    primaryClass = "astro-ph.CO",
    month = "11",
    year = "2025"
}

@article{DESI:2025ffm,
    author = "Ahlen, S. P. and others",
    collaboration = "DESI",
    title = "{Positive Neutrino Masses with DESI DR2 via Matter Conversion to Dark Energy}",
    eprint = "2504.20338",
    archivePrefix = "arXiv",
    primaryClass = "astro-ph.CO",
    reportNumber = "FERMILAB-PUB-25-0288-PPD",
    doi = "10.1103/yb2k-kn7h",
    journal = "Phys. Rev. Lett.",
    volume = "135",
    number = "8",
    pages = "081003",
    year = "2025"
}

@article{PhysRevD.64.112007,
  title = {Evidence for neutrino oscillations from the observation of ${\overline{\ensuremath{\nu}}}_{e}$ appearance in a ${\overline{\ensuremath{\nu}}}_{\ensuremath{\mu}}$ beam},
  author = {Aguilar, A. and Auerbach, L. B. and Burman, R. L. and Caldwell, D. O. and Church, E. D. and Cochran, A. K. and Donahue, J. B. and Fazely, A. and Garvey, G. T. and Gunasingha, R. M. and Imlay, R. and Louis, W. C. and Majkic, R. and Malik, A. and Metcalf, W. and Mills, G. B. and Sandberg, V. and Smith, D. and Stancu, I. and Sung, M. and Tayloe, R. and VanDalen, G. J. and Vernon, W. and Wadia, N. and White, D. H. and Yellin, S.},
  collaboration = {LSND Collaboration},
  journal = {Phys. Rev. D},
  volume = {64},
  issue = {11},
  pages = {112007},
  numpages = {22},
  year = {2001},
  month = {Nov},
  publisher = {American Physical Society},
  doi = {10.1103/PhysRevD.64.112007},
  url = {https://link.aps.org/doi/10.1103/PhysRevD.64.112007}
}

@article{Babu:1991at,
    author = "Babu, K. S. and Rothstein, I. Z.",
    title = "{Relaxing nucleosynthesis bounds on sterile-neutrinos}",
    reportNumber = "MDDP-PP-91-290",
    doi = "10.1016/0370-2693(92)90860-7",
    journal = "Phys. Lett. B",
    volume = "275",
    pages = "112--118",
    year = "1992"
}

@article{Gariazzo:2019gyi,
    author = "Gariazzo, S. and de Salas, P. F. and Pastor, S.",
    title = "{Thermalisation of sterile neutrinos in the early Universe in the 3+1 scheme with full mixing matrix}",
    eprint = "1905.11290",
    archivePrefix = "arXiv",
    primaryClass = "astro-ph.CO",
    doi = "10.1088/1475-7516/2019/07/014",
    journal = "JCAP",
    volume = "07",
    pages = "014",
    year = "2019"
}

@article{Dasgupta:2013zpn,
    author = "Dasgupta, Basudeb and Kopp, Joachim",
    title = "{Cosmologically Safe eV-Scale Sterile Neutrinos and Improved Dark Matter Structure}",
    eprint = "1310.6337",
    archivePrefix = "arXiv",
    primaryClass = "hep-ph",
    doi = "10.1103/PhysRevLett.112.031803",
    journal = "Phys. Rev. Lett.",
    volume = "112",
    number = "3",
    pages = "031803",
    year = "2014"
}

@article{Hannestad:2013ana,
    author = "Hannestad, Steen and Hansen, Rasmus Sloth and Tram, Thomas",
    title = "{How Self-Interactions can Reconcile Sterile Neutrinos with Cosmology}",
    eprint = "1310.5926",
    archivePrefix = "arXiv",
    primaryClass = "astro-ph.CO",
    doi = "10.1103/PhysRevLett.112.031802",
    journal = "Phys. Rev. Lett.",
    volume = "112",
    number = "3",
    pages = "031802",
    year = "2014"
}

@article{Chu:2018gxk,
    author = "Chu, Xiaoyong and Dasgupta, Basudeb and Dentler, Mona and Kopp, Joachim and Saviano, Ninetta",
    title = "{Sterile neutrinos with secret interactions{\textemdash}cosmological discord?}",
    eprint = "1806.10629",
    archivePrefix = "arXiv",
    primaryClass = "hep-ph",
    reportNumber = "TIFR/TH/18-17",
    doi = "10.1088/1475-7516/2018/11/049",
    journal = "JCAP",
    volume = "11",
    pages = "049",
    year = "2018"
}

@article{Archidiacono:2014nda,
    author = "Archidiacono, Maria and Hannestad, Steen and Hansen, Rasmus Sloth and Tram, Thomas",
    title = "{Cosmology with self-interacting sterile neutrinos and dark matter - A pseudoscalar model}",
    eprint = "1404.5915",
    archivePrefix = "arXiv",
    primaryClass = "astro-ph.CO",
    doi = "10.1103/PhysRevD.91.065021",
    journal = "Phys. Rev. D",
    volume = "91",
    number = "6",
    pages = "065021",
    year = "2015"
}

@article{Archidiacono:2020yey,
    author = "Archidiacono, Maria and Gariazzo, Stefano and Giunti, Carlo and Hannestad, Steen and Tram, Thomas",
    title = "{Sterile neutrino self-interactions: $H_0$ tension and short-baseline anomalies}",
    eprint = "2006.12885",
    archivePrefix = "arXiv",
    primaryClass = "astro-ph.CO",
    doi = "10.1088/1475-7516/2020/12/029",
    journal = "JCAP",
    volume = "12",
    pages = "029",
    year = "2020"
}

@article{Ivanez-Ballesteros:2024nws,
    author = "Iv{\'a}{\~n}ez-Ballesteros, Pilar and Volpe, M. Cristina",
    title = "{Constraints on neutrino-Majoron coupling using SN1987A data}",
    eprint = "2410.11517",
    archivePrefix = "arXiv",
    primaryClass = "hep-ph",
    doi = "10.1103/d4vp-m261",
    journal = "Phys. Rev. D",
    volume = "112",
    number = "10",
    pages = "L101301",
    year = "2025"
}

@article{Chang:2022aas,
    author = "Chang, Po-Wen and Esteban, Ivan and Beacom, John F. and Thompson, Todd A. and Hirata, Christopher M.",
    title = "{Toward Powerful Probes of Neutrino Self-Interactions in Supernovae}",
    eprint = "2206.12426",
    archivePrefix = "arXiv",
    primaryClass = "hep-ph",
    doi = "10.1103/PhysRevLett.131.071002",
    journal = "Phys. Rev. Lett.",
    volume = "131",
    number = "7",
    pages = "071002",
    year = "2023"
}

@article{Shalgar:2019rqe,
    author = "Shalgar, Shashank and Tamborra, Irene and Bustamante, Mauricio",
    title = "{Core-collapse supernovae stymie secret neutrino interactions}",
    eprint = "1912.09115",
    archivePrefix = "arXiv",
    primaryClass = "astro-ph.HE",
    doi = "10.1103/PhysRevD.103.123008",
    journal = "Phys. Rev. D",
    volume = "103",
    number = "12",
    pages = "123008",
    year = "2021"
}

@article{Das:2017iuj,
    author = "Das, Anirban and Dighe, Amol and Sen, Manibrata",
    title = "{New effects of non-standard self-interactions of neutrinos in a supernova}",
    eprint = "1705.00468",
    archivePrefix = "arXiv",
    primaryClass = "hep-ph",
    reportNumber = "TIFR-TH-17-18",
    doi = "10.1088/1475-7516/2017/05/051",
    journal = "JCAP",
    volume = "05",
    pages = "051",
    year = "2017"
}

@article{Farzan:2002wx,
    author = "Farzan, Yasaman",
    title = "{Bounds on the coupling of the Majoron to light neutrinos from supernova cooling}",
    eprint = "hep-ph/0211375",
    archivePrefix = "arXiv",
    reportNumber = "SLAC-PUB-9543, SISSA-69-2002-EP",
    doi = "10.1103/PhysRevD.67.073015",
    journal = "Phys. Rev. D",
    volume = "67",
    pages = "073015",
    year = "2003"
}

@article{Bustamante:2020mep,
    author = "Bustamante, Mauricio and Rosenstr{\o}m, Charlotte and Shalgar, Shashank and Tamborra, Irene",
    title = "{Bounds on secret neutrino interactions from high-energy astrophysical neutrinos}",
    eprint = "2001.04994",
    archivePrefix = "arXiv",
    primaryClass = "astro-ph.HE",
    doi = "10.1103/PhysRevD.101.123024",
    journal = "Phys. Rev. D",
    volume = "101",
    number = "12",
    pages = "123024",
    year = "2020"
}

@article{Esteban:2021tub,
    author = "Esteban, Ivan and Pandey, Sujata and Brdar, Vedran and Beacom, John F.",
    title = "{Probing secret interactions of astrophysical neutrinos in the high-statistics era}",
    eprint = "2107.13568",
    archivePrefix = "arXiv",
    primaryClass = "hep-ph",
    reportNumber = "FERMILAB-PUB-21-328-T, nuhep-th/21-06",
    doi = "10.1103/PhysRevD.104.123014",
    journal = "Phys. Rev. D",
    volume = "104",
    number = "12",
    pages = "123014",
    year = "2021"
}

@article{Ioka:2014kca,
    author = "Ioka, Kunihto and Murase, Kohta",
    title = "{IceCube PeV{\textendash}EeV neutrinos and secret interactions of neutrinos}",
    eprint = "1404.2279",
    archivePrefix = "arXiv",
    primaryClass = "astro-ph.HE",
    reportNumber = "KEK-TH-1723, KEK-COSMO-141",
    doi = "10.1093/ptep/ptu090",
    journal = "PTEP",
    volume = "2014",
    number = "6",
    pages = "061E01",
    year = "2014"
}

@article{Agostini:2015nwa,
    author = "Agostini, M. and others",
    title = "{Results on $\beta \beta $ decay with emission of two neutrinos or Majorons in$^{76}$ Ge from GERDA Phase I}",
    eprint = "1501.02345",
    archivePrefix = "arXiv",
    primaryClass = "nucl-ex",
    doi = "10.1140/epjc/s10052-015-3627-y",
    journal = "Eur. Phys. J. C",
    volume = "75",
    number = "9",
    pages = "416",
    year = "2015"
}

@article{Blum:2018ljv,
    author = "Blum, Kfir and Nir, Yosef and Shavit, Michal",
    title = "{Neutrinoless double-beta decay with massive scalar emission}",
    eprint = "1802.08019",
    archivePrefix = "arXiv",
    primaryClass = "hep-ph",
    doi = "10.1016/j.physletb.2018.08.022",
    journal = "Phys. Lett. B",
    volume = "785",
    pages = "354--361",
    year = "2018"
}

@article{Pasquini:2015fjv,
    author = "Pasquini, P. S. and Peres, O. L. G.",
    title = "{Bounds on Neutrino-Scalar Yukawa Coupling}",
    eprint = "1511.01811",
    archivePrefix = "arXiv",
    primaryClass = "hep-ph",
    doi = "10.1103/PhysRevD.93.053007",
    journal = "Phys. Rev. D",
    volume = "93",
    number = "5",
    pages = "053007",
    year = "2016",
    note = "[Erratum: Phys.Rev.D 93, 079902 (2016)]"
}

@article{Dev:2024ygx,
    author = "Dev, P. S. Bhupal and Kim, Doojin and Sathyan, Deepak and Sinha, Kuver and Zhang, Yongchao",
    title = "{New laboratory constraints on neutrinophilic mediators}",
    eprint = "2407.12738",
    archivePrefix = "arXiv",
    primaryClass = "hep-ph",
    reportNumber = "CETUP-2024-005",
    doi = "10.1016/j.physletb.2025.139765",
    journal = "Phys. Lett. B",
    volume = "868",
    pages = "139765",
    year = "2025"
}

@article{Brdar:2020nbj,
    author = "Brdar, Vedran and Lindner, Manfred and Vogl, Stefan and Xu, Xun-Jie",
    title = "{Revisiting neutrino self-interaction constraints from $Z$ and $\tau$ decays}",
    eprint = "2003.05339",
    archivePrefix = "arXiv",
    primaryClass = "hep-ph",
    doi = "10.1103/PhysRevD.101.115001",
    journal = "Phys. Rev. D",
    volume = "101",
    number = "11",
    pages = "115001",
    year = "2020"
}

@article{Brune:2018sab,
    author = {Brune, Tim and P{\"a}s, Heinrich},
    title = "{Massive Majorons and constraints on the Majoron-neutrino coupling}",
    eprint = "1808.08158",
    archivePrefix = "arXiv",
    primaryClass = "hep-ph",
    reportNumber = "DO-TH 18/23",
    doi = "10.1103/PhysRevD.99.096005",
    journal = "Phys. Rev. D",
    volume = "99",
    number = "9",
    pages = "096005",
    year = "2019"
}

@article{Lyu:2020lps,
    author = "Lyu, Kun-Feng and Stamou, Emmanuel and Wang, Lian-Tao",
    title = "{Self-interacting neutrinos: Solution to Hubble tension versus experimental constraints}",
    eprint = "2004.10868",
    archivePrefix = "arXiv",
    primaryClass = "hep-ph",
    doi = "10.1103/PhysRevD.103.015004",
    journal = "Phys. Rev. D",
    volume = "103",
    number = "1",
    pages = "015004",
    year = "2021"
}

@article{Blinov:2019gcj,
    author = "Blinov, Nikita and Kelly, Kevin James and Krnjaic, Gordan Z and McDermott, Samuel D",
    title = "{Constraining the Self-Interacting Neutrino Interpretation of the Hubble Tension}",
    eprint = "1905.02727",
    archivePrefix = "arXiv",
    primaryClass = "astro-ph.CO",
    reportNumber = "FERMILAB-PUB-19-175-A-T",
    doi = "10.1103/PhysRevLett.123.191102",
    journal = "Phys. Rev. Lett.",
    volume = "123",
    number = "19",
    pages = "191102",
    year = "2019"
}

@article{Cyr-Racine:2013jua,
    author = "Cyr-Racine, Francis-Yan and Sigurdson, Kris",
    title = "{Limits on Neutrino-Neutrino Scattering in the Early Universe}",
    eprint = "1306.1536",
    archivePrefix = "arXiv",
    primaryClass = "astro-ph.CO",
    doi = "10.1103/PhysRevD.90.123533",
    journal = "Phys. Rev. D",
    volume = "90",
    number = "12",
    pages = "123533",
    year = "2014"
}

@article{Kreisch:2019yzn,
    author = "Kreisch, Christina D. and Cyr-Racine, Francis-Yan and Dor{\'e}, Olivier",
    title = "{Neutrino puzzle: Anomalies, interactions, and cosmological tensions}",
    eprint = "1902.00534",
    archivePrefix = "arXiv",
    primaryClass = "astro-ph.CO",
    doi = "10.1103/PhysRevD.101.123505",
    journal = "Phys. Rev. D",
    volume = "101",
    number = "12",
    pages = "123505",
    year = "2020"
}

@article{Brinckmann:2020bcn,
    author = "Brinckmann, Thejs and Chang, Jae Hyeok and LoVerde, Marilena",
    title = "{Self-interacting neutrinos, the Hubble parameter tension, and the cosmic microwave background}",
    eprint = "2012.11830",
    archivePrefix = "arXiv",
    primaryClass = "astro-ph.CO",
    reportNumber = "YITP-SB-2020-40",
    doi = "10.1103/PhysRevD.104.063523",
    journal = "Phys. Rev. D",
    volume = "104",
    number = "6",
    pages = "063523",
    year = "2021"
}

@article{Das:2020xke,
    author = "Das, Anirban and Ghosh, Subhajit",
    title = "{Flavor-specific interaction favors strong neutrino self-coupling in the early universe}",
    eprint = "2011.12315",
    archivePrefix = "arXiv",
    primaryClass = "astro-ph.CO",
    reportNumber = "SLAC-PUB-17547",
    doi = "10.1088/1475-7516/2021/07/038",
    journal = "JCAP",
    volume = "07",
    pages = "038",
    year = "2021"
}

@article{Archidiacono:2013dua,
    author = "Archidiacono, Maria and Hannestad, Steen",
    title = "{Updated constraints on non-standard neutrino interactions from Planck}",
    eprint = "1311.3873",
    archivePrefix = "arXiv",
    primaryClass = "astro-ph.CO",
    doi = "10.1088/1475-7516/2014/07/046",
    journal = "JCAP",
    volume = "07",
    pages = "046",
    year = "2014"
}

@article{Forastieri:2019cuf,
    author = "Forastieri, Francesco and Lattanzi, Massimiliano and Natoli, Paolo",
    title = "{Cosmological constraints on neutrino self-interactions with a light mediator}",
    eprint = "1904.07810",
    archivePrefix = "arXiv",
    primaryClass = "astro-ph.CO",
    doi = "10.1103/PhysRevD.100.103526",
    journal = "Phys. Rev. D",
    volume = "100",
    number = "10",
    pages = "103526",
    year = "2019"
}

@article{Poudou:2025qcx,
    author = "Poudou, Ad{\`e}le and Simon, Th{\'e}o and Montandon, Thomas and Teixeira, Elsa M. and Poulin, Vivian",
    title = "{Self-interacting neutrinos in light of recent CMB and LSS data}",
    eprint = "2503.10485",
    archivePrefix = "arXiv",
    primaryClass = "astro-ph.CO",
    doi = "10.1103/mljb-42fm",
    journal = "Phys. Rev. D",
    volume = "112",
    number = "10",
    pages = "103535",
    year = "2025"
}

@article{AtacamaCosmologyTelescope:2025nti,
    author = "Calabrese, Erminia and others",
    collaboration = "Atacama Cosmology Telescope",
    title = "{The Atacama Cosmology Telescope: DR6 constraints on extended cosmological models}",
    eprint = "2503.14454",
    archivePrefix = "arXiv",
    primaryClass = "astro-ph.CO",
    reportNumber = "FERMILAB-PUB-25-0157-PPD",
    doi = "10.1088/1475-7516/2025/11/063",
    journal = "JCAP",
    volume = "11",
    pages = "063",
    year = "2025"
}

@article{Camarena:2024daj,
    author = "Camarena, David and Cyr-Racine, Francis-Yan",
    title = "{Strong constraints on a simple self-interacting neutrino cosmology}",
    eprint = "2403.05496",
    archivePrefix = "arXiv",
    primaryClass = "astro-ph.CO",
    doi = "10.1103/PhysRevD.111.023504",
    journal = "Phys. Rev. D",
    volume = "111",
    number = "2",
    pages = "023504",
    year = "2025"
}

@article{Beacom:2004yd,
    author = "Beacom, John F. and Bell, Nicole F. and Dodelson, Scott",
    title = "{Neutrinoless universe}",
    eprint = "astro-ph/0404585",
    archivePrefix = "arXiv",
    reportNumber = "FERMILAB-PUB-04-050-A",
    doi = "10.1103/PhysRevLett.93.121302",
    journal = "Phys. Rev. Lett.",
    volume = "93",
    pages = "121302",
    year = "2004"
}

@article{Barenboim:2020vrr,
    author = "Barenboim, Gabriela and Chen, Joe Zhiyu and Hannestad, Steen and Oldengott, Isabel M. and Tram, Thomas and Wong, Yvonne Y. Y.",
    title = "{Invisible neutrino decay in precision cosmology}",
    eprint = "2011.01502",
    archivePrefix = "arXiv",
    primaryClass = "astro-ph.CO",
    doi = "10.1088/1475-7516/2021/03/087",
    journal = "JCAP",
    volume = "03",
    pages = "087",
    year = "2021"
}

@article{Archidiacono:2015oma,
    author = "Archidiacono, Maria and Hannestad, Steen and Hansen, Rasmus Sloth and Tram, Thomas",
    title = "{Sterile neutrinos with pseudoscalar self-interactions and cosmology}",
    eprint = "1508.02504",
    archivePrefix = "arXiv",
    primaryClass = "astro-ph.CO",
    doi = "10.1103/PhysRevD.93.045004",
    journal = "Phys. Rev. D",
    volume = "93",
    number = "4",
    pages = "045004",
    year = "2016"
}

@article{Bashinsky:2003tk,
    author = "Bashinsky, Sergei and Seljak, Uros",
    title = "{Neutrino perturbations in CMB anisotropy and matter clustering}",
    eprint = "astro-ph/0310198",
    archivePrefix = "arXiv",
    doi = "10.1103/PhysRevD.69.083002",
    journal = "Phys. Rev. D",
    volume = "69",
    pages = "083002",
    year = "2004"
}

@article{Baumann:2015rya,
    author = "Baumann, Daniel and Green, Daniel and Meyers, Joel and Wallisch, Benjamin",
    title = "{Phases of New Physics in the CMB}",
    eprint = "1508.06342",
    archivePrefix = "arXiv",
    primaryClass = "astro-ph.CO",
    doi = "10.1088/1475-7516/2016/01/007",
    journal = "JCAP",
    volume = "01",
    pages = "007",
    year = "2016"
}

@article{Pan:2016zla,
    author = "Pan, Zhen and Knox, Lloyd and Mulroe, Brigid and Narimani, Ali",
    title = "{Cosmic Microwave Background Acoustic Peak Locations}",
    eprint = "1603.03091",
    archivePrefix = "arXiv",
    primaryClass = "astro-ph.CO",
    doi = "10.1093/mnras/stw833",
    journal = "Mon. Not. Roy. Astron. Soc.",
    volume = "459",
    number = "3",
    pages = "2513--2524",
    year = "2016"
}

@article{Follin:2015hya,
    author = "Follin, Brent and Knox, Lloyd and Millea, Marius and Pan, Zhen",
    title = "{First Detection of the Acoustic Oscillation Phase Shift Expected from the Cosmic Neutrino Background}",
    eprint = "1503.07863",
    archivePrefix = "arXiv",
    primaryClass = "astro-ph.CO",
    doi = "10.1103/PhysRevLett.115.091301",
    journal = "Phys. Rev. Lett.",
    volume = "115",
    number = "9",
    pages = "091301",
    year = "2015"
}

@article{Mangano:2001iu,
    author = "Mangano, G. and Miele, G. and Pastor, S. and Peloso, M.",
    title = "{A Precision calculation of the effective number of cosmological neutrinos}",
    eprint = "astro-ph/0111408",
    archivePrefix = "arXiv",
    reportNumber = "DSF-37-2001, MPI-PHT-2001-51",
    doi = "10.1016/S0370-2693(02)01622-2",
    journal = "Phys. Lett. B",
    volume = "534",
    pages = "8--16",
    year = "2002"
}

@article{Froustey:2020mcq,
    author = "Froustey, Julien and Pitrou, Cyril and Volpe, Maria Cristina",
    title = "{Neutrino decoupling including flavour oscillations and primordial nucleosynthesis}",
    eprint = "2008.01074",
    archivePrefix = "arXiv",
    primaryClass = "hep-ph",
    doi = "10.1088/1475-7516/2020/12/015",
    journal = "JCAP",
    volume = "12",
    pages = "015",
    year = "2020"
}

@article{Bennett:2020zkv,
    author = "Bennett, Jack J. and Buldgen, Gilles and De Salas, Pablo F. and Drewes, Marco and Gariazzo, Stefano and Pastor, Sergio and Wong, Yvonne Y. Y.",
    title = "{Towards a precision calculation of $N_{\rm eff}$ in the Standard Model II: Neutrino decoupling in the presence of flavour oscillations and finite-temperature QED}",
    eprint = "2012.02726",
    archivePrefix = "arXiv",
    primaryClass = "hep-ph",
    reportNumber = "CPPC-2020-10",
    doi = "10.1088/1475-7516/2021/04/073",
    journal = "JCAP",
    volume = "04",
    pages = "073",
    year = "2021"
}

@article{Drewes:2024wbw,
    author = "Drewes, Marco and Georis, Yannis and Klasen, Michael and Wiggering, Luca Paolo and Wong, Yvonne Y. Y.",
    title = "{Towards a precision calculation of N $_{eff}$ in the Standard Model. Part III. Improved estimate of NLO contributions to the collision integral}",
    eprint = "2402.18481",
    archivePrefix = "arXiv",
    primaryClass = "hep-ph",
    reportNumber = "CPPC-2024-01, MS-TP-24-06",
    doi = "10.1088/1475-7516/2024/06/032",
    journal = "JCAP",
    volume = "06",
    pages = "032",
    year = "2024"
}

@article{Hou:2011ec,
    author = "Hou, Zhen and Keisler, Ryan and Knox, Lloyd and Millea, Marius and Reichardt, Christian",
    title = "{How Massless Neutrinos Affect the Cosmic Microwave Background Damping Tail}",
    eprint = "1104.2333",
    archivePrefix = "arXiv",
    primaryClass = "astro-ph.CO",
    doi = "10.1103/PhysRevD.87.083008",
    journal = "Phys. Rev. D",
    volume = "87",
    pages = "083008",
    year = "2013"
}

@misc{Cherry:2014xra,
    author = "Cherry, John F. and Friedland, Alexander and Shoemaker, Ian M.",
    title = "{Neutrino Portal Dark Matter: From Dwarf Galaxies to IceCube}",
    eprint = "1411.1071",
    archivePrefix = "arXiv",
    primaryClass = "hep-ph",
    reportNumber = "CP3-Origins-2014-034, DIAS-2014-34, LA-UR-14-28339",
    month = "11",
    year = "2014"
}

@article{Aloni:2023tff,
    author = "Aloni, Daniel and Joseph, Melissa and Schmaltz, Martin and Weiner, Neal",
    title = "{Dark Radiation from Neutrino Mixing after Big Bang Nucleosynthesis}",
    eprint = "2301.10792",
    archivePrefix = "arXiv",
    primaryClass = "astro-ph.CO",
    doi = "10.1103/PhysRevLett.131.221001",
    journal = "Phys. Rev. Lett.",
    volume = "131",
    number = "22",
    pages = "221001",
    year = "2023"
}

@article{Das:2025asx,
    author = "Das, Anirban and Dev, P. S. Bhupal and Gao, Christina and Ghosh, Subhajit and Kim, Taegyun",
    title = "{Impostor among Neutrinos: Dark Radiation Masquerading as Self-Interacting Neutrinos}",
    eprint = "2506.08085",
    archivePrefix = "arXiv",
    primaryClass = "hep-ph",
    reportNumber = "MITP-25-044",
    doi = "10.1103/jprg-jll6",
    journal = "Phys. Rev. Lett.",
    volume = "136",
    number = "13",
    pages = "131003",
    year = "2026"
}

@ARTICLE{2011JCAP...07..034B,
       author = {{Blas}, Diego and {Lesgourgues}, Julien and {Tram}, Thomas},
        title = "{The Cosmic Linear Anisotropy Solving System (CLASS). Part II: Approximation schemes}",
      journal = {\jcap},
         year = 2011,
        month = jul,
       volume = {2011},
       number = {7},
          eid = {034},
        pages = {034},
          doi = {10.1088/1475-7516/2011/07/034},
archivePrefix = {arXiv},
       eprint = {1104.2933},
 primaryClass = {astro-ph.CO},
       adsurl = {https://ui.adsabs.harvard.edu/abs/2011JCAP...07..034B}
}

@ARTICLE{2011arXiv1104.2932L,
       author = {{Lesgourgues}, Julien},
        title = "{The Cosmic Linear Anisotropy Solving System (CLASS) I: Overview}",
      journal = {arXiv e-prints},
         year = 2011,
        month = apr,
          eid = {arXiv:1104.2932},
        pages = {arXiv:1104.2932},
          doi = {10.48550/arXiv.1104.2932},
archivePrefix = {arXiv},
       eprint = {1104.2932},
 primaryClass = {astro-ph.IM},
       adsurl = {https://ui.adsabs.harvard.edu/abs/2011arXiv1104.2932L}
}

\appendix

\section*{Appendix}

\section*{Additional figures and tables}
\begin{figure}[!h]
    \centering
    \includegraphics[width=\linewidth]{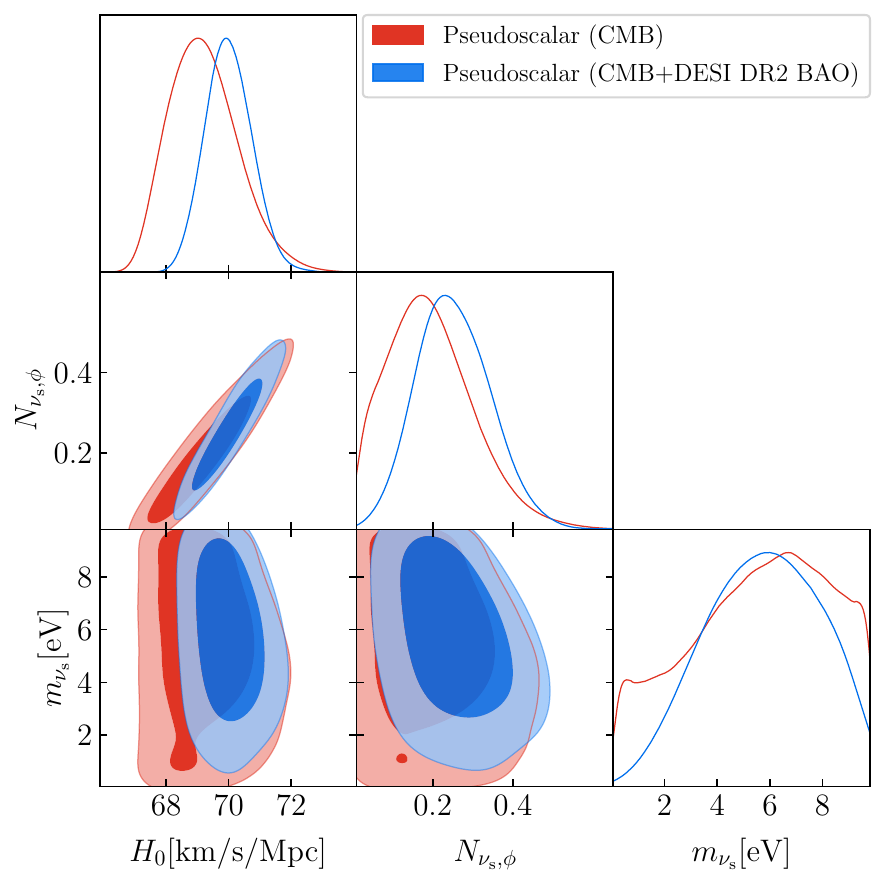}
    \caption{Marginalised 68\% and 95\% contours and 1D posteriors for the sterile neutrino and pseudoscalar parameters ($\nsp$ and $\ms$) and for $H_0$, obtained by fitting CMB alone (in red) and in combination with DESI DR2 BAO (in blue).
    }
    \label{fig:triangle}
\end{figure}

\begin{figure}[!h]
    \centering  \includegraphics[width=\linewidth]{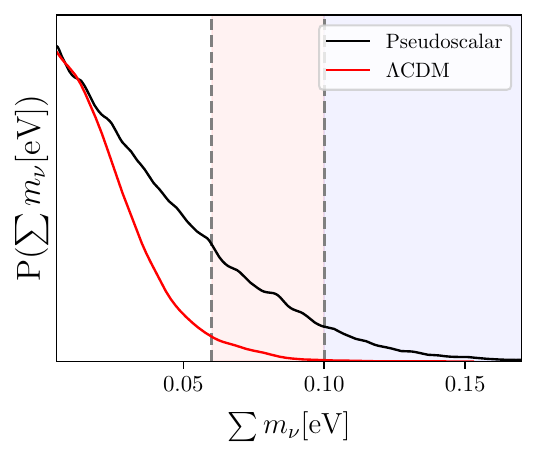}
    \caption{1D posterior on the sum of active neutrino masses in the pseudoscalar model (solid black line) and in $\Lambda$CDM (solid red line) obtained by fitting CMB+DESI DR2 BAO data. The dashed grey vertical lines show the minimum $\summnu$ in normal (pink shade) and in inverted (blue shade) ordering}
    \label{fig:mnu}
\end{figure}

Table \ref{tab:params_main} and Fig.~\ref{fig:triangle} show the constraints obtained in $\Lambda$CDM and in our pseudoscalar model with the baseline CMB datset alone or in combination with DESI DR2 BAO. Fig.~\ref{fig:mnu} shows the constraints on the summed active neutrino mass in the same models with the combined dataset. 
We refer the reader to the main text for further discussion of the results.
\begin{table*}
\centering
\begin{tabular}{|l|cc|cc|}
\hline
 & \multicolumn{2}{c}{CMB} & \multicolumn{2}{c|}{CMB+DESI DR2 BAO} \\
\cline{2-3} \cline{4-5}
Parameter 
& $\Lambda$CDM & Pseudo-Scalar 
& $\Lambda$CDM & Pseudo-Scalar \\
\hline
{\boldmath $H_0$ [km/s/Mpc]}                      & $67.37 \pm 0.37$ &$69.2 \pm 1.1$ &$68.12 \pm 0.26$ &$70.01 \pm 0.74$\\
{\boldmath$\Omega_\mathrm{b} h^2$} &$0.02220 \pm 0.00011$  &$0.02231 \pm 0.00013$ &$0.02229 \pm 0.00010$&$0.02237 \pm 0.00012$  \\
{\boldmath$\Omega_\mathrm{c} h^2$} & $0.11954 \pm 0.00086$  & $0.1230 \pm 0.0022$&$0.11785 \pm 0.00059$&$0.1236 \pm 0.0022$ \\
{\boldmath$\ln (10^{10} A_{\rm s})$}     &$3.054 \pm 0.011$  & $3.059 \pm 0.011$ &$3.066 \pm 0.012$&$3.063 \pm 0.011$\\
{\boldmath$n_{\rm s}$}                   & $0.9647 \pm 0.0035$ & $0.9720 \pm 0.0060$&$0.9688 \pm 0.0032$&$0.9762 \pm 0.0049$ \\
{\boldmath $\tau_{\rm reio}$}                      &$0.0595 \pm 0.0057$  & $0.0609 \pm 0.0061$&$0.0664 \pm 0.0065$ &$0.0632 \pm 0.0061$ \\

\hline
{\boldmath $\nsp$}                      & $-$ & $ < 0.40$ &$-$&$0.252 \pm 0.093$ \\
{\boldmath $\ms$}                      & $-$ & n.c. &$-$&$>2$ \\

\hline
$\Omega_\mathrm{m}$ &$0.3139 \pm 0.0051$  & $0.3050 \pm 0.0070$&$0.3033 \pm 0.0034$ &$0.2991 \pm 0.0037$\\
$r_{\rm drag}$ & $147.41 \pm 0.20$ &$145.2 \pm 1.3$&$147.76 \pm 0.17$ &$144.7 \pm 1.1$ \\
$\sigma_{8}$&$0.8171 \pm 0.0038$&$0.8247 \pm 0.0066$&$0.8169 \pm 0.0040$&$0.8279 \pm 0.0065$\\
$S_{8}$&${0.8358}^{+0.0074}_{-0.0081}$&$0.8315 \pm 0.0086$&$0.8214 \pm 0.0058$&${0.8266}^{+0.0067}_{-0.0070}$\\

\hline
$\chi^2_{\rm min}$ & 11853.3 &11850.44 & 11870.68&11861.7 \\
\hline
\end{tabular}
\caption{\label{tab:params_main}
Mean values and 68\% credible intervals for the cosmological parameters and for selected derived parameters in $\Lambda$CDM and in the pseudoscalar model, obtained either with the baseline CMB dataset alone or in combination with DESI DR2 BAO. We also report the minimum $\chi^2$ values at the best fit.}
\end{table*}

\section*{Including ACT DR6}
\label{sec:act}
We incorporate ACT DR6 primary anisotropy results using the lite version of ACT  TT,TE,EE data in the multipole range $600<l<8500$  \cite{AtacamaCosmologyTelescope:2025blo}. We then cut Planck data at $l>1000$ in TT and $l>600$ in TE,EE, as suggested in \cite{AtacamaCosmologyTelescope:2025blo}.
When adding ACT DR6 to our baseline CMB dataset, the posterior of $\nsp$ peaks in zero with $\nsp<0.13$ (95\% CL), while $\ms$ remains unconstrained. The overall $\chi^2$ does not improve relative to $\Lambda$CDM, $\Delta \chi^2=-0.01$. The possibility to solve the Hubble tension is also reduced, since in the pseudoscalar model we get $H_0=68.23 \pm 0.56$, while with the same data the $\Lambda$CDM model gives $H_0=67.73 \pm 0.37$. Our findings are consistent with the ACT DR6 analysis showing a preference for $\neff<3.044$ \cite{AtacamaCosmologyTelescope:2025nti}.

\section*{Equilibration}
\label{sec:equilibration}

\begin{figure}[!h]
    \centering \includegraphics[width=\linewidth]{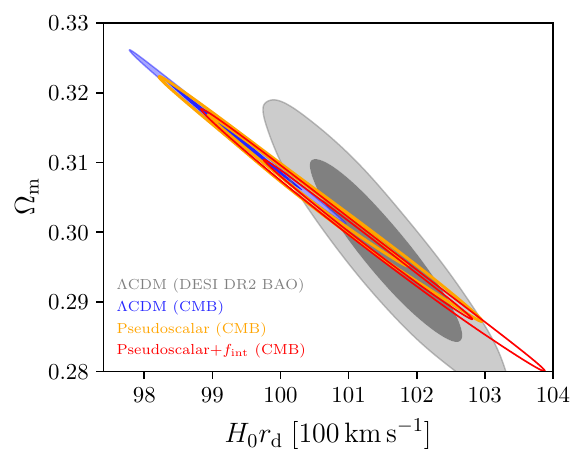}
    \caption{68\% and 95\% marginalized contours on the $\Omega_{m}$, $H_{0}r_{\rm d}$ plane obtained by fitting CMB data alone in $\Lambda$CDM (blue filled contours) and in the pseudoscalar model without (orange empty contours, as in Fig.~\ref{fig:bestfit_bao}) and with (red empty contours) a varying equilibration fraction. The grey filled contours are obtained by fitting only DESI DR2 data in $\Lambda$CDM.} 
    \label{fig:equi}
\end{figure}

After recoupling between sterile neutrinos and pseudoscalars, oscillations and collisions may in principle equilibrate sterile and active neutrinos \cite{Cherry:2014xra, Aloni:2023tff, Das:2025asx}. However, for $g \sim 3 \times 10^{-6}$ the effective interaction rate $\Gamma_{\rm eff}=\frac{1}{2} \sin^2(2 \theta_{\rm m}) \Gamma_{\rm s}$ exceeds the expansion rate at $T \lesssim 10$ eV, close to the onset of sterile neutrino annihilations. As a results, complete equilibration is never achieved. Although flavour conversion becomes efficient at late times, sterile neutrinos simultaneously become nonrelativistic and annihilate into pseudoscalars, preventing the establishment of a common thermal bath.
\begin{figure}[!h]
    \centering
\includegraphics[width=\linewidth]{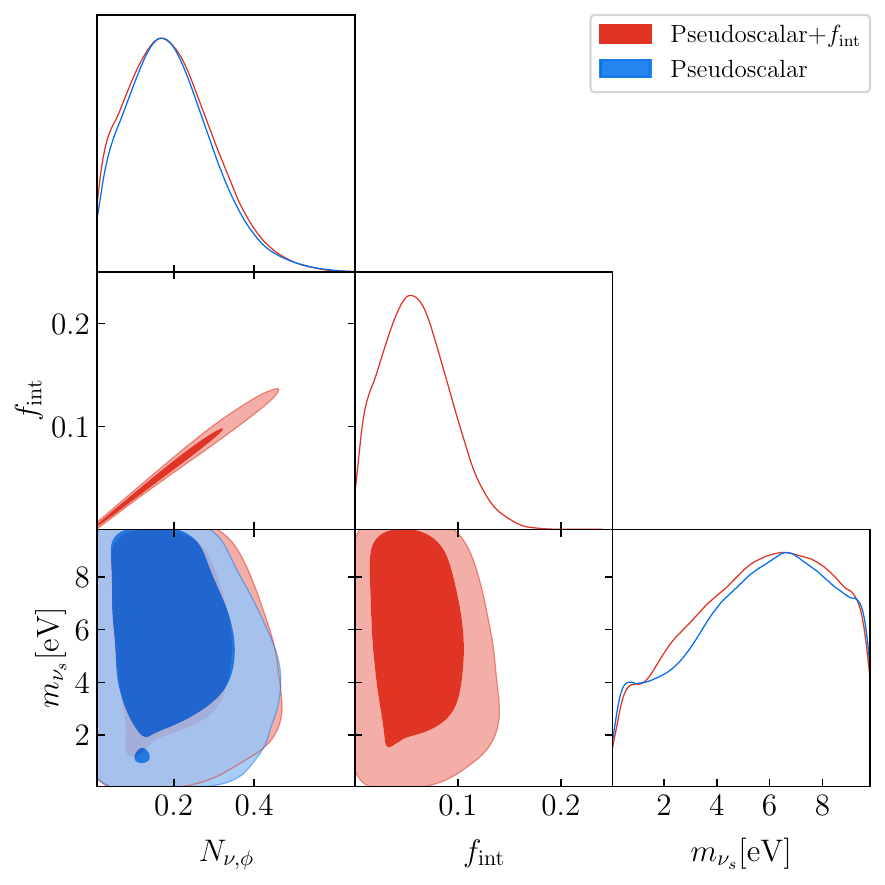}
    \caption{Marginalised 68\% and 95\% contours and 1D posteriors for  $\nsp$, $\ms$, and $f_{\rm int}$ obtained by fitting CMB alone in the pseudoscalar model, with (red) and without (blue) partial equilibratio.}
\label{fig:pseudo_fraction}
\end{figure}
Nevertheless, here we investigate the impact of a partial equilibration by varying the fraction of $\neff$ behaving as a fluid with pseudoscalar interactions.
We vary the total $\neff$ and the fraction $f_{\rm int}$, defining $\nsp=f_{\rm int}\neff$, while the free streaming component is $(1-f_{\rm int})\neff$. In this case we assume that all neutrino species are massless. 
Using CMB only data, we find $\neff=3.20 \pm 0.17$ and $f_{c}<0.12$ (95\% CL).
(see fig \ref{fig:pseudo_fraction}).
This confirms that current CMB data prefer active neutrinos to be mainly free streaming and only the extra component to be self-interacting. Interestingly, the consistency between CMB and DESI DR2 BAO improves when allowing for a varying fraction of the interacting sterile neutrinos (see Fig.~\ref{fig:equi}). The improvement in goodness of fit is $\Delta \chi^2=-3.60$ ($\Delta \chi^2=-10.64$) relative to $\Lambda$CDM from CMB alone (from CMB + DESI DR2 BAO). The credible interval for the Hubble parameter becomes $H_0=69.6 \pm 1.4 \, {\rm km}/{\rm s}/{\rm Mpc}$, reducing the tension with SH0ES to the 2.1$\sigma$ level.

\end{document}